\documentclass[reqno,aps,prd,twocolumn,showpacs,showkeys,preprintnumbers,amsmath,amssymb,amsfonts]{revtex4}
\newcommand{\wh}{\widehat}
\usepackage{graphics}
\usepackage[dvips]{graphicx}% Include figure files
\usepackage{dcolumn}% Align table columns on decimal point
\usepackage{bm}% bold math
\usepackage[bookmarksnumbered]{hyperref}
\usepackage{color}

\numberwithin{equation}{section}

\begin{document}

\preprint{SPR1.71}

\pdfbookmark{Periodic relativity: deflection of light, acceleration, rotation curves}{tit}
\title{Periodic relativity: deflection of light, acceleration, rotation curves}% Force line breaks with \\
%\homepage[]{http://groups.msn.com/PeriodicRelativity/welcome1.msnw}


\author{Vikram H. Zaveri}
%\email{cons\_eng1@yahoo.com}

\email{zaverivik@hotmail.com}

\affiliation{B-4/6, Avanti Apt., Harbanslal Marg, Sion, Mumbai 400022 INDIA}

%\thanks{}

%\noaffiliation{}
Journal: Progress in Physics, 2015, v.11(1), 43--49.\\
%Published online: Nov. 15, 2009.\\
%http://ptep-online.com/index_files/2015/PP-40-11.PDF
%DOI: \href{http://dx.doi.org/10.1007/s10714-009-0908-5}{10.1007/s10714-009-0908-5}
%\date{Sept. 5, 2014}

\begin{abstract}
Vectorial analysis relating to derivation of deflection of light is presented. Curvilinear acceleration is distinguished from the Newtonian polar conic acceleration. The difference between the two is due to the curvature term. Lorentz invariant expression for acceleration is derived. A physical theory of rotation curves of galaxies based on second solution to Einstein's field equation is presented. Theory is applied to Milky Way, M31, NGC3198 and Solar system. Modified Kepler's third law yields correct orbital periods of stars in a galaxy. Deviation factor in the line element of the theory happens to be the ratio of the Newtonian gravitational acceleration to the measured acceleration of the star in the galaxy. Therefore this deviation factor can replace the MOND function.
\end{abstract}

\pacs{04.50.Kd,\: 98.35.Df,\: 98.62.Dm,\: 04.20.Cv}
\keywords{Two-body problem, Acceleration, Bending of light, Rotation curves.}

\maketitle
\section{Introduction}
The article presented here is only a small element of a much larger formulation \cite{44,77,20,50,104,55} proposed
to arrive at a theory of quantum gravity and cosmology. Physicists have put in considerable efforts to unify general relativity and quantum mechanics but without success. The string theory and loop quantum gravity are still far from their goal. 

Scientists are looking for a unified theory of creation. To achieve this objective, the physicists have set up two principal goals. First is the search for the fundamental building block of the universe. Second is the unification of four fundamental forces in nature. This constitutes the mainstream physics. The theory presented here regards these two principal goals as speculative and not plausible and hence the deviation from the mainstream physics. 

Another feature of the mainstream physics is that most of the physicists if not all, consider consciousness \cite{104,55} as something outside the domain of physics and therefore when they talk about theory of everything, they really mean theory of everything excluding consciousness. As per the current understanding in the physical and life sciences, much of the scientific literature maintain strict distinction between consciousness and matter. The former is considered sentient and the later insentient. Many people are of the opinion that the existence of consciousness in this universe is a reality and the big bang theory could not be considered complete till it can account for the presence of consciousness along with the other forms of insentient matter. 

Having rejected the two principal goals of the mainstream physics, this theory proposes that everything in the universe is reducible to energy. Therefore unity behind four forces (bosons), fermions and leptons should be sought in energy. Another point this theory makes is that the consciousness and energy are two states of one and the same thing which you may call the fundamental substance (Spirit) of the universe. Fundamental building block of the universe is assumed to be a micro entity, but the fundamental substance of the universe is all pervasive and ever remains undivided. 

In this theory space and time does not have any physical existence, but they exist only in the human mind as imaginary artifacts. Comparatively, the energy has some real existence and it is found in myriads of forms. Again the energy is always associated with oscillations and motion, without exception. When these oscillation and motion of the energy subside, it gets transformed into the unmanifest which is not the energy and therefore does not gravitate. This unmanifest is motionless without any oscillations and therefore impossible to detect like empty space.

The idea of space-time arise in the human mind by way of delusion. When a particle wave is presented to a physicist, instead of seeing the oscillating energy, what he does is, superimposes the idea of wavelength and period on this wave and sees the space-time. All the geometrical theories in physics are founded upon such delusion. In periodic quantum gravity (PQG), the time does not flow in one direction, but one gets the sense of time by comparing one period of time with another. Hence time is a periodic phenomenon and periods are inverse of frequencies. Therefore in PQG, the Hubble parameter is associated with the frequency of the particle. Both have the same units. This eliminates the problem of time which plagues the Wheeler De Witt equation and its associated theories like loop quantum gravity, Hartle-Hawking wavefunction of the universe etc.

Advantage of Periodic relativity (PR) over general relativity can be seen in its use of revised principle of equivalence which states that the gravitational mass is equal to the relativistic mass. Application of this principle gives a very simple derivation for the orbital period derivative of the binary star \cite{20}. And most important of all, allows the unification of periodic relativity with quantum mechanics. Because of this revised principle of equivalence, (modified) Newton's inverse square law of gravitation can be merged with the (modified) Schrodinger Wave equation which gives the basis for periodic quantum gravity and cosmology theory \cite{50}. PR satisfies Einstein's field equations but does not utilize weak field approximation.

The reason general relativity (GR) got plagued with these two problems (the problem of time associated with Wheeler De Witt equation and the inaccurate notion that the gravitational mass is equal to the inertial mass) is its dependence on the weak field approximation. The use of weak field approximation automatically locks the theory into having these two problems. When you depend on weak field approximation, you cannot treat time as a periodic phenomenon and you cannot introduce energy momentum invariant into Newton's inverse square law.

Another problem with GR is that the universe in this theory begins with a mixture of energy (radiation) and matter field. It doesn't even bother to explain where these two things come from. Another contradiction is that the equivalence of mass and energy is the biggest feature of GR at the same time they must have the universe begin with a mixture of energy (radiation) and the matter field. And all the physicists find it very comfortable to ignore the presence of life and consciousness in the universe. At the same time they must have a theory of everything. 

Periodic quantum gravity and cosmology \cite{50} is based on the idea that there is a connection between consciousness and energy \cite{104}. Based on these ideas PQG proposes a unified field of consciousness (UFC) \cite{55} underlying the entire universe from which comes the energy and matter fields of the big bang theory. In relating the consciousness and the energy the periodic nature of the time is the most essential factor. You don't need any clock operators of the Wheeler De Witt theory.

On the quantum mechanical side I don't think Dirac's linear representation of the wave function is very accurate because spin in that theory is not a part of the dynamics of motion but it is introduced as a perturbation just like in Darwin and Pauli theories. Also, the selection of the radial momentum operator is somewhat arbitrary and it isn't Hermitian as pointed out by several authors. These deficiencies are removed in the modified Schrodinger wave equation \cite{77} in which spin is directly introduced in the Laplacian operator. This gives exactly same energy levels for hydrogen atom as in Dirac's theory and also it's application to heavy quarkonium spectra gives data which are spin dependent. 

When these two theories, the periodic relativity and the relativistic wave mechanics are united, the result is the periodic quantum gravity and cosmology theory \cite{50} which yields the entire table of standard model particles from a single formula. There is no other theory of quantum gravity that can do this. 

Current article presents some corrections in previous article \cite{44} and perfects the derivation for the deflection of light. It develops Lorentz invariant expression for the acceleration and provides solution for the rotation curves of galaxies which does not exist in GR. This solution does not have a discontinuity like the one in the MOND function. The transition from short distances to astronomical distances is continuous. This theory gives perfect fit for the rotation curves which MOND theory cannot give.

\section{Curvilinear Gravity}
In the earlier article "Periodic relativity: basic framework of the theory" \cite{44}, we obtained correct deflection of light in Newtonian theory by multiplying both sides of Newton's inverse square law of gravitation by the factor $\left(\cos{\psi}+\sin{\psi}\right)$. As shown in Figs.1 and 2 of that article, $\psi$ is the angle between the radial vector and the tangential velocity vector. Explanation given below makes it more clear that the theory is Lorentz invariant and factor $\left(\cos{\psi}+\sin{\psi}\right)$ introduces geodesic like trajectories. The details are as follows. After very elaborate analysis, we arrive at Newton's inverse square law given by
\begin{align}\label{1.1}%4.63
m_0\frac{d^2\mathbf{r}}{dt^2}=
-\frac{GM_0m_0}{r^2}\mathbf{\hat r}.
\end{align}
where $GM_0=\mu$. Here we introduce the dynamic weak equivalence principle which states that the gravitation mass is equal to the relativistic mass. Therefore Eq.~\eqref{1.1} becomes
\begin{align}\label{1.2}%4.64
m\frac{d^2\mathbf{r}}{dt^2}=
-\frac{\mu m}{r^2}\mathbf{\hat r}.
\end{align}
In classical mechanics, we have two different expressions for the acceleration acting on a body in motion. One is a general expression $d\mathbf{v}/dt$ in cartesian coordinates which include the curvature term, and another is for Newtonian gravity in polar coordinates $d^2\mathbf{r}/dt^2$ based on the angular momentum vector $\mathbf{h}$, which is supposed to be a constant in order to satisfy Kepler's third law of equal areas in equal times. In periodic relativity \cite{44} we have shown that these two accelerations are not equal. At the same time we have maintained that the velocity vectors in both coordinate systems are equal, $\mathbf{v}=d\mathbf{r}/dt$. The reason for this is that the Newtonian gravity ignores the variation of angle $\psi$ along the trajectory by assuming constant $\mathbf{h}$.

\begin{figure*}[ht]
\includegraphics[width=12cm]{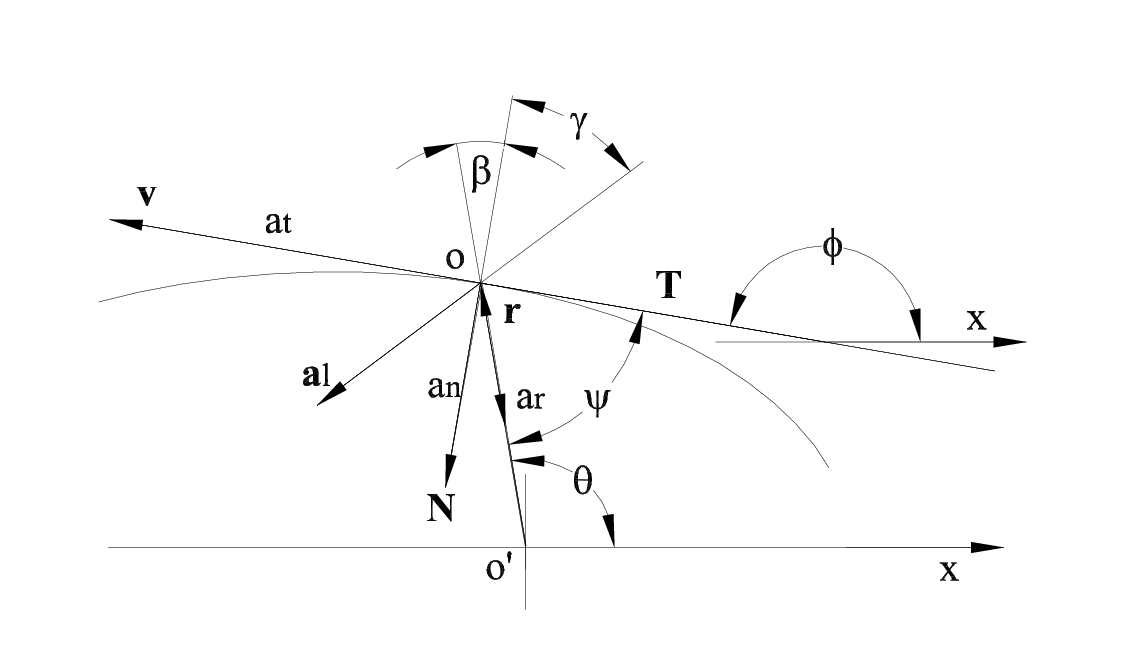}
\caption[]{Vectors in a two-body system\label{Fig.1}}
\end{figure*}

As shown in Fig.~\ref{Fig.1}, this angle $\psi$ is related to curvature through the expression 
\begin{align}\label{1.3}%4.64
\phi=\theta+\psi.
\end{align}
where $d\phi/ds=\kappa$ is the curvature. Newtonian gravity ignores this curvature term by assuming constant $\psi=\pi/2$. This can be verified from following arguments.
\begin{align}\label{1.4}%4.40a
\mathbf{h}=\frac{\mathbf{L}}{m}=\frac{\mathbf{p}\boldsymbol{\times}\mathbf{r}}{m}\equiv\frac{|\mathbf{p}||\mathbf{r}|\sin{\psi}}{m}\mathbf{\hat h}=r^2\frac{d\theta}{dt}\sin{\psi}\mathbf{\hat h}.
\end{align}
From Eq.~\eqref{1.4} we can see that $\mathbf{h}$ can be the desired constant only if $\sin{\psi}=1$. This shows that the very foundation of Newtonian gravity ignores the curvature of the trajectory of the orbiting body. Hence in periodic relativity it is considered unreasonable to equate the cartesian acceleration $d\mathbf{v}/dt$ with the Newtonian polar acceleration $d^2\mathbf{r}/dt^2$.

In order to account for the variation of angle $\psi$ along the trajectory, we propose that the absolute sum of vector and scalar products of $(\mu/r^2)\mathbf{\hat r}$ and $\mathbf{\hat a}$ is equal to magnitude of $d\mathbf{v}/dt$. The relation of these vectors to angle $\psi$ is shown in Fig.~\ref{Fig.1}.

\begin{align}\label{1.5}%4.65
\left|\frac{d\mathbf{v}}{dt}\right|=\left|-\left|\mathbf{\hat a}\boldsymbol{\times}\frac{\mu}{r^2}\mathbf{\hat r}\right|-\frac{\mu}{r^2}\mathbf{\hat r}\cdot\mathbf{\hat a}\right|.
\end{align}
\begin{align}\label{1.6}
\left|\frac{d\mathbf{v}}{dt}\right|=\left||\mathbf{\hat a}|\left|\frac{\mu}{r^2}\mathbf{\hat r}\right|\sin{(\beta+\gamma)}\mathbf{\hat h}\right|+\left|\frac{\mu}{r^2}\mathbf{\hat r}\right||\mathbf{\hat a}|\cos{(\beta+\gamma)}.
\end{align}
where
\begin{align}\label{1.61}
\beta=\left(\frac{\pi}{2}-\psi\right).
\end{align}
\begin{align}\label{1.7}
\gamma=\tan^{-1}\left(\frac{a_t}{a_n}\right).
\end{align}
Various magnitudes of the parameters shown in Fig.~\ref{Fig.1} are as follows.
\begin{align}\label{1.8}
\mathbf{a}_l=\frac{d\mathbf{v}}{dt}.
\end{align}
\begin{align}\label{1.9}
a_t=\left(\frac{d^2s}{dt^2}+\frac{v}{\nu}\frac{d\nu}{dt}\right).
\end{align}
\begin{align}\label{1.10}
a_n=\kappa\left(\frac{ds}{dt}\right)^2.
\end{align}
\begin{align}\label{1.11}
a_r=-\frac{\mu}{r^2}=\left| \frac{d^2\mathbf{r}}{dt^2} \right|.
\end{align}
\begin{align}\label{1.12}
\mathbf{v}=\frac{d\mathbf{r}}{dt}.
\end{align}
Substitution of Eq.~\eqref{1.61} in Eq.~\eqref{1.6} gives
\begin{align}\label{1.13}
\left|\frac{d\mathbf{v}}{dt}\right|=\frac{\mu}{r^2}\left(\cos{(\psi-\gamma)}+\sin{(\psi-\gamma)}\right).
\end{align}
When the tangential component of the acceleration is absent then we have $a_t\mathbf{\wh T}=0$. This gives $\gamma=0$ and Eq.~\eqref{1.13} reduces to
\begin{align}\label{1.14}
\left|\frac{d\mathbf{v}}{dt}\right|=\frac{\mu}{r^2}\left(\cos{\psi}+\sin{\psi}\right).
\end{align}
Similarly we can show that
\begin{align}\label{1.15}
\left|\frac{d\mathbf{v}}{dt}\right|=\left| \frac{d^2\mathbf{r}}{dt^2} \right| \left(\cos{(\psi-\gamma)}+\sin{(\psi-\gamma)}\right).
\end{align}
The first term on the right of Eq.~\eqref{1.13} can be interpreted as an angular acceleration vector with its axis perpendicular to the plane of motion. This could be the additional acceleration quantity responsible for the rotation of the velocity vector $\mathbf{v}$ about the coordinate origin $o$, causing the curvature of the trajectory. 

\subsection{Lorentz invariant acceleration}
Little diversion here. In the earlier work \cite{44}, we introduced deviation to the flat Minkowski metric due to the gravitational field in the form,
\begin{align}\label{1.16}%c1
\left(\frac{dt}{d\tau}\right)^2=\gamma^{2n}=(1-\beta^2)^{-n},
\end{align}
Here I propose a correction to our theory and change the method of introducing the deviation so that the deviation factor $n$ is directly introduced in the Lorentz transformation equation as given below.
\begin{align}\label{1.17}%c2
\left(\frac{d\tau}{dt}\right)^2=(1-n\beta^2),
\end{align}
where $t$ is the coordinate time, $\tau$ the proper time of the orbiting body, $n$ is a real number and $\beta=v/c$. The corresponding line element in polar coordinates is,
\begin{align}\label{1.18}%c3
ds^2=c^2dt^2-ndr^2-nr^2d\theta^2-n(r^2\sin^2{\theta})d\phi^2.
\end{align}
We showed \cite{44} that the line element Eq.~\eqref{1.18} satisfies Einstein's field equations for any constant value of $n$. For any constant value of $n$, metric~\eqref{1.18} always remain flat. This is similar to the line element in Friedmann model when curvature factor $K=0$. The change made in equation~\eqref{1.17} does not alter any of the previous derivations. 

Coming back to the main topic, in relativity we can either write our equations in terms of proper time or alternatively we can write them in terms of relativistic mass. Eq.~\eqref{1.17} can be written as 
\begin{align}\label{1.19}
\left(\frac{d\tau}{dt}\right)^2=(1-n\beta^2)=\left(\frac{m_0}{m}\right)^2=\left(\frac{E_0}{E}\right)^2,
\end{align}
where $E=mc^2=h\nu$. This gives
\begin{align}\label{1.20}
E=(E_0^2+nE^2\beta^2)^{1/2}.
\end{align}
Differentiating w.r.t. time we get
\begin{align}\label{1.21}
\frac{dE}{dt}=\mathbf{\hat v}F=n\left(m\mathbf{a}+\frac{h\mathbf{v}}{c^2}\frac{d\nu}{dt}\right).
\end{align}
Here we arrive at the same relation that we described as true force in the previous article \cite{44} except that now we have introduced the deviation factor $n$. I like to further point out a correction that this true force is same as the Lorentz force. Here we have used the relation $E=mc^2=h\nu$. Therefore 
\begin{align}\label{1.22}%3.13
\mathbf{F}=\frac{d\mathbf{p}}{dt}=\frac{dm\mathbf{v}}{dt}=n\left(m\mathbf{a}+\frac{h\mathbf{v}}{c^2}\frac{d\nu}{dt}\right),	
\end{align}
where $\mathbf{F}$ is the Lorentz force and $\mathbf{v}$ the velocity vector and $\mathbf{a}$
is the classical acceleration of the particle given by
\begin{align}\label{1.23}
\mathbf{a}=\left(\frac{d^2s}{dt^2}\mathbf{\wh T}+\kappa\left(\frac{ds}{dt}\right)^2\mathbf{\wh N}\right).
\end{align}
Therefore,\\
Lorentz force = Classical force + de Broglie force.\\
From Eq.~\eqref{1.22} we can define Lorentz invariant acceleration $\mathbf{a}_l$ as
\begin{align}\label{1.24}
n\mathbf{a}_l=n\left(\left(\frac{d^2s}{dt^2}+\frac{v}{\nu}\frac{d\nu}{dt}\right)\mathbf{\wh T}+\kappa\left(\frac{ds}{dt}\right)^2\mathbf{\wh N}\right).
\end{align}
The de Broglie force acts along the tangent vector.
Now we equate Lorentz force with the gravitational force given by Eq.~\eqref{1.13}
\begin{align}\label{1.25}
\begin{split}
\left|nm\mathbf{a}_l\right|&=\left|m\frac{d\mathbf{v}}{dt}\right|=nm\left(\left(\frac{d^2s}{dt^2}+\frac{v}{\nu}\frac{d\nu}{dt}\right)\mathbf{\wh T}+\kappa\left(\frac{ds}{dt}\right)^2\mathbf{\wh N}\right)\\&=
\frac{\mu m}{r^2}\left(\cos{(\psi-\gamma)}+\sin{(\psi-\gamma)}\right).
\end{split}
\end{align}
\begin{align}\label{1.26}
\begin{split}
\left|\mathbf{a}_l\right|&=\left|\frac{1}{n}\frac{d\mathbf{v}}{dt}\right|=\left(\left(\frac{d^2s}{dt^2}+\frac{v}{\nu}\frac{d\nu}{dt}\right)\mathbf{\wh T}+\kappa\left(\frac{ds}{dt}\right)^2\mathbf{\wh N}\right)\\&=
\frac{\mu}{nr^2}\left(\cos{(\psi-\gamma)}+\sin{(\psi-\gamma)}\right).
\end{split}
\end{align}

\subsection{Bending of light in periodic relativity}
For the bending of light around the sun, we introduce light parameters $v=ds/dt=c$, $d^2s/dt^2 =0$ and $cdt=ds$, along with $\kappa=d\phi/ds$ for the curvature of the trajectory in Eq.~\eqref{1.26}. In this case we will have $d\nu/dt=0$ because the ray is equally blue shifted and then red shifted, and the frequency shift is $0$ at the limb of the sun.
This gives,
\begin{align}\label{4.10}
\left|\frac{c^2}{\nu}\frac{d\nu}{ds}\mathbf{\wh T}+c^2\frac{d\phi}{ds}\mathbf{\wh N}\right|=\frac{\mu}{nr^2}\left(\cos{(\psi-\gamma)}+\sin{(\psi-\gamma)}\right).
\end{align}
Multiplying both sides by $d\psi$, we get
\begin{align}\label{4.11}
\begin{split}
&\left|\frac{1}{\nu} d\nu d\psi \mathbf{\wh T}+ d\phi d\psi \mathbf{\wh N}\right|\\&=\frac{\mu}{nc^2r^2}\left(\cos{(\psi-\gamma)}+\sin{(\psi-\gamma)}\right)ds d\psi.
\end{split}
\end{align}
We integrate both sides with proper limits. For the star light approaching the sun we get,
\begin{align}\label{4.12}
\begin{split}
&\left|\int_{\nu_1}^{\nu_2} \int_\pi^{\frac{\pi}{2}}\frac{1}{\nu}d\nu d\psi \mathbf{\wh T}+\int_{-\phi}^0 \int_\pi^{\frac{\pi}{2}}d\phi d\psi \mathbf{\wh N}\right|\\&=\frac{\mu}{nc^2}\int_{-\infty}^0 
\int_\pi^{\frac{\pi}{2}} \frac{1}{r^2}\left(\cos{(\psi-\gamma)}+\sin{(\psi-\gamma)}\right)d\psi ds.
\end{split}
\end{align}
For the star light approaching earth from the limb of the sun we get,
\begin{align}\label{4.13}
\begin{split}
&\left|\int_{\nu_2}^{\nu_1} \int_{\frac{\pi}{2}}^0\frac{1}{\nu}d\nu d\psi \mathbf{\wh T}+\int_0^{-\phi} \int_{\frac{\pi}{2}}^0 d\phi d\psi \mathbf{\wh N}\right|\\&=\frac{\mu}{nc^2}\int_0^{\infty} 
\int_{\frac{\pi}{2}}^0 \frac{1}{r^2}\left(\cos{(\psi-\gamma)}+\sin{(\psi-\gamma)}\right)d\psi ds.
\end{split}
\end{align}
\begin{align}\label{4.14}
\begin{split}
&\left|(\text{ln}\nu_2-\text{ln}\nu_1)\mathbf{\wh T}+\phi \mathbf{\wh N}\right|\\&=\frac{\mu}{nc^2}\int_{-\infty}^0 
\int_\pi^{\frac{\pi}{2}} \frac{1}{r^2}\left(\cos{(\psi-\gamma)}+\sin{(\psi-\gamma)}\right)d\psi ds.
\end{split}
\end{align}
\begin{align}\label{4.15}
\begin{split}
&\left|(\text{ln}\nu_1-\text{ln}\nu_2)\mathbf{\wh T}+\phi \mathbf{\wh N}\right|\\&=\frac{\mu}{nc^2}\int_0^{\infty} 
\int_{\frac{\pi}{2}}^0 \frac{1}{r^2}\left(\cos{(\psi-\gamma)}+\sin{(\psi-\gamma)}\right)d\psi ds.
\end{split}
\end{align}
If we add l.h.s. of Eqs.~\eqref{4.14}\ and~\eqref{4.15} we get,
\begin{align}\label{4.16}
l.h.s. = \left|0.\mathbf{\wh T}+2\phi \mathbf{\wh N}\right|.
\end{align}
From Eq.~\eqref{4.16} we see that the magnitude of the tangential component is zero. Therefore $\gamma=0$. Hence substituting $r^2=s^2+\Delta^2$ in Eqs.~\eqref{4.14} and~\eqref{4.15} we get
\begin{align}\label{4.17}
2\phi=\frac{4\mu}{nc^2\Delta}.
\end{align}
It is obvious from Eq.~\eqref{4.17} that the value of constant $n$ is $1$ and not 0 as was assumed in earlier article \cite{44}. $n=1$ corresponds to the flat Minkowski metric therefore both the bending of light and the gravitational frequency shift can be explained corresponding to $n=1$. Not only that, but no matter what gets measured in future experiments such as LATOR, the new measurement can easily be made to fit Eq.~\eqref{4.17} by adjusting the constant $n$.

\subsection{Curvic and conic gravity}
Newtonian gravity is based on the constant vector $\mathbf{h}$ which yields the conic sections. Therefore we can distinguish the gravity that uses the Lorentz invariant acceleration as the curvilinear (or curvic) gravity and the Newtonian gravity with constant $\mathbf{h}$ as the conic gravity. Accelerations of the curvic and conic gravity are related by Eq.~\eqref{1.15}.

%\begin{widetext}
\begin{table}
	 \caption{Milky Way rotation curve based on proper time.\label{tab:Table1}}
 \begin{ruledtabular}
		\begin{tabular} {llllll}
%\hline	
		$r (kpc)$ &$v (km/s)$ &$k\times 10^{-81}$ &$\hspace*{7 mm}n$ &$d\tau/dt$ \\  \hline 
		$7.5$ &$216$ &$1.79546$ &$0.62593$ &$1-1.6246\times 10^{-7}$\\
		$8.0$ &$220$	&$2.10050$ &$0.56566$ &$1-1.5231\times 10^{-7}$\\
		$12.5$ &$227$ &$7,52624$ &$0.34004$ &$1-9.748\times 10^{-8}$\\
		$17.5$ &$179$  &$33.2129$ &$0.39061$ &$1-6.9628\times 10^{-8}$\\
		$22.5$ &$168$ &$80.1362$ &$0.34490$ &$1-5.4155\times 10^{-8}$\\
		$27.5$ &$183$ &$123.309$ &$0.23782$ &$1-4.43091\times 10^{-8}$\\
		$32.5$ &$143$ &$333.332$ &$0.32956$ &$1-3.7492\times 10^{-8}$\\
		$37.5$ &$170$ &$362.322$ &$0.20210$ &$1-3.2493\times 10^{-8}$\\
		$42.5$ &$183$ &$455.160$ &$0.15388$ &$1-2.8670\times 10^{-8}$\\
		$47.5$ &$165$ &$781.650$ &$0.16936$ &$1-2.5652\times 10^{-8}$\\
		$55$ &$183$ &$986.474$ &$0.11891$ &$1-2.2154\times 10^{-8}$\\
%\hline
				\end{tabular}
 \end{ruledtabular}
\end{table}
%\end{widetext}  

%\begin{widetext}
\begin{table}
	 \caption{Solar system rotation curve based on proper time.\label{tab:Table2}}
 \begin{ruledtabular}
		\begin{tabular} {llllll}
%\hline	
Planet	&$r\times10^{-9}(m)$ &$v (km/s)$ &$k$ &$\hspace*{3 mm}n$ \\  \hline 
Mercury	&$57.91$ &$47.87$ &$1.12\times 10^{43}$ &$1.000103$ \\
Venus		&$108.21$ &$35.02$	&$1.37\times 10^{44}$ &$1.000059$ \\
Earth		&$149.6$ &$29.78$ &$5.01\times 10^{44}$ &$1.000332$ \\
Mars		&$227.92$ &$24.13$  &$2.69\times 10^{45}$ &$1.000065$ \\
Jupiter	&$778.57$ &$13.07$ &$3.66\times 10^{47}$ &$0.997876$ \\
Saturn	&$1433.53$ &$9.69$ &$4.16\times 10^{48}$ &$0.985986$ \\
Uranus	&$2872.46$ &$6.81$ &$6.78\times 10^{49}$ &$0.99627$ \\
Neptune	&$4495.06$ &$5.43$ &$4.08\times 10^{50}$ &$1.00136$ \\
Pluto		&$5869.66$ &$4.72$ &$1.20\times 10^{51}$ &$1.014912$ \\
Moon		&$0.3844$ &$1.023$ &$2.16\times 10^{34}$ &$0.990824$ \\
%\hline
				\end{tabular}
 \end{ruledtabular}
\end{table}
%\end{widetext}  

It also needs to be understood that $d^2\mathbf{r}/dt^2$ is a radial vector but $d\mathbf{r}/dt$ is not a radial vector which acts along the velocity vector $\mathbf{v}$. Moreover, the constant vector $\mathbf{h}$ does not play any role in defining the velocity vector $\mathbf{v}$. Therefore factor $\left(\cos{\psi}+\sin{\psi}\right)$ does not appear in this expression of velocity $\mathbf{v}=d\mathbf{r}/dt$ which remains unaltered. This can be verified from following analysis. By definition we have
\begin{align}\label{1.27}
\cos{\psi}=\frac{dr}{ds},\ \ \ \text{and}\ \ \   \sin{\psi}=\frac{rd\theta}{ds}.
\end{align}
\begin{align}\label{1.28}
\frac{d\mathbf{r}}{dt}=
\left(\frac{dr}{dt}\mathbf{\hat r}+\frac{rd\theta}{dt}\boldsymbol{\hat \theta}\right).
\end{align}
\begin{align}\label{1.29}
\frac{d\mathbf{r}}{dt}=
\frac{ds}{dt}\left(\cos{(\psi+\theta)}\mathbf{i}+\sin{(\psi+\theta)}\mathbf{j}\right).
\end{align}
Substitution of Eq.~\eqref{1.3} gives
\begin{align}\label{1.30}
\frac{d\mathbf{r}}{dt}=
\frac{ds}{dt}\sqrt{\left(\cos^2{\phi}+\sin^2{\phi}\right)}\mathbf{\wh T}=\frac{ds}{dt}\mathbf{\wh T}=\mathbf{v}.
\end{align}
From Fig.~\ref{Fig.1} we can verify that the unit vector acting at an angle $\phi$ is $\mathbf{\wh T}$. Therefore 
Eq.~\eqref{1.30} is not influenced by the constant $\mathbf{h}$ assumption.

\section{Rotation curves of galaxies}
Earlier \cite{44} we obtained two solutions to Einstein's field equations,
\begin{align}\label{4.58a}
\left(\frac{r}{n}\frac{\partial n}{\partial r}\right)=0 \qquad and \qquad \left(\frac{r}{n}\frac{\partial n}{\partial r}\right)=-4.
\end{align}
So far we have seen the application of the first solution which requires $n$ to be any real number constant. Now we look at the second solution which we can write as
\begin{align}\label{1.31}
\int\frac{\partial n}{n}=-4\int\frac{\partial r}{r}.
\end{align}
\begin{align}\label{1.32}
\text{ln}(nr^4)=C.
\end{align}
where $C$ is a constant of integration. This gives
\begin{align}\label{1.33}
n=\frac{e^C}{r^4}=\frac{k}{r^4}.
\end{align}
\begin{align}\label{1.34}
\text{where}\ \ \  k=e^C=\text{constant}.
\end{align}
In this second solution $n$ need not be a constant. We make use of Eq.~\eqref{1.26} in order to apply the second solution to rotation curves of a galaxies. Assuming circular orbit we substitute $\psi=\pi/2$ and $\gamma=0$. This gives
\begin{align}\label{1.35}
|\mathbf{a}|=\frac{\mu}{nr^2}=\frac{\mu r^2}{k}=\frac{v^2}{r}.
\end{align}
\begin{align}\label{1.36}
k=\frac{\mu r^3}{v^2}.
\end{align}
We can write Eq.~\eqref{1.35} as
\begin{align}\label{1.37}
v^2=\frac{4\pi^2r^2}{P^2}=\frac{\mu}{nr}.
\end{align}
\begin{align}\label{1.37a}
P=\frac{2 \pi r}{v}.
\end{align}
\begin{align}\label{1.38}
P^2=\frac{4\pi^2r^3n}{\mu}.
\end{align}
For $n=1$, Eq.~\eqref{1.38} reduces to Kepler's third law, where $P$ is the orbital period.  
Substituting Eq.~\eqref{1.36} in Eq.~\eqref{1.33} and Eq.~\eqref{1.17} we can compute the ratio $d\tau/dt$. We can apply these equations of stellar motion to Blue Horizontal-Branch (BHB) halo stars of the Milky Way \cite{19}. The circular velocity estimates are based on Naab's simulation \cite{41}. To this data, one additional data point for solar radius of $8kpc$ \cite{26} is added and the results obtained from Eqs.~\eqref{1.36},~\eqref{1.33} and~\eqref{1.17} are shown in Table~\ref{tab:Table1}. Computed values are based on the stellar mass at the galactic center, which is $5.0924\times10^{10}M_\odot$ \cite{18,42}. Observed values of $r$ and circular velocities constrain the integration constant $k$ which provides a measure of non-uniform distribution of the galactic matter and the cold dark matter at a given radius. Hence it is appropriate to describe $k$ as a galactic matter distribution constant. We also find that Eqs.~\eqref{1.37a} and~\eqref{1.38} both yield exactly the same orbital period when velocity and deviation $n$ along with the galactic stellar mass are used from the Tables. For the Sun, both yield 223.4 million years.
%\begin{widetext}
\begin{table}
	 \caption{M31 rotation curve. $k$ in m$^4$, $P$ in yrs.\label{tab:Table3}}
 \begin{ruledtabular}
		\begin{tabular} {llllll}
%\hline	
		$r (kpc)$ &$v (km/s)$ &$k\times 10^{-81}$ &$\hspace*{1 mm}n$ &$d\tau/dt$ &$P\times 10^{-8}$ \\  \hline 
		$8.5$ &$232.0$ &$6.23$ &$1.316$ &$1-3.94\times 10^{-7}$ &$2.250$\\
		$12.5$ &$251.2$	&$16.89$ &$0.763$ &$1-2.68\times 10^{-7}$ &$3.057$\\
		$16.5$ &$251.6$ &$38.74$ &$0.576$ &$1-2.03\times 10^{-7}$ &$4.029$\\
		$20.5$ &$227.4$  &$90.94$ &$0.568$ &$1-1.63\times 10^{-7}$ &$5.538$\\
		$24.5$ &$226.2$ &$156.89$ &$0.480$ &$1-1.367\times 10^{-7}$ &$6.654$\\
		$28.5$ &$218.8$ &$263.96$ &$0.441$ &$1-1.175\times 10^{-7}$ &$8.0$\\
		$32.5$ &$224.7$ &$371.15$ &$0.367$ &$1-1.030\times 10^{-7}$ &$8.885$\\
		$36.5$ &$240.1$ &$460.47$ &$0.286$ &$1-9.178\times 10^{-8}$ &$9.339$\\
%\hline
				\end{tabular}
 \end{ruledtabular}
\end{table}
%\end{widetext}  

%\begin{widetext}
\begin{table}
	 \caption{NGC3198 rotation curve. $k$ in m$^4$, $P$ in yrs.\label{tab:Table4}}
 \begin{ruledtabular}
		\begin{tabular} {llllll}
%\hline	
$r (kpc)$ &$v (km/s)$ &$k\times 10^{-79}$ &$\hspace*{1 mm}n$ &$d\tau/dt$ &$P\times 10^{-8}$ \\  \hline 
		$0.68$ &$55$ &$0.202$ &$10.45$ &$1-1.76\times 10^{-7}$ &$0.759$\\
		$1.36$ &$92$	&$0.579$ &$1.868$ &$1-8.79\times 10^{-8}$ &$0.908$\\
		$2.72$ &$123$ &$2.593$ &$0.522$ &$1-4.39\times 10^{-8}$ &$1.358$\\
		$5.44$ &$147$  &$14.52$ &$0.183$ &$1-2.2\times 10^{-8}$ &$2.273$\\
		$8.16$ &$156$ &$43.52$ &$0.108$ &$1-1.466\times 10^{-8}$ &$3.213$\\
		$13.6$ &$154$ &$206.78$ &$0.066$ &$1-8.79\times 10^{-9}$ &$5.425$\\
		$19.04$ &$148$ &$614.36$ &$0.0515$ &$1-6.28\times 10^{-9}$ &$7.903$\\
		$24.48$ &$148$ &$1305.7$ &$0.040$ &$1-4.88\times 10^{-9}$ &$10.16$\\
		$29.92$ &$149$ &$2352.1$ &$0.0323$ &$1-3.99\times 10^{-9}$ &$12.33$\\%\hline
				\end{tabular}
 \end{ruledtabular}
\end{table}
%\end{widetext}  

Table~\ref{tab:Table2} shows solar system data from NASA planet fact sheets. Radial distance equal to semi major axis and mean orbital velocity are used. $k$ and $n$ are computed using Eqs.~\eqref{1.36} and~\eqref{1.33}. $(1-d\tau/dt)$ are of order $10^{-8}$ to $10^{-12}$ and not shown in the table. In case of moon, earth mass $5.9736\times10^{24}$ Kg. is used. Value of $n$ for Mercury shown in Table~\ref{tab:Table2} should not be compared with that used in the derivation of perihelic precession \cite{44} because here we have used second solution of Einstein's field equations with constant $k$, where as perihelic precession is derived from the first solution of Einstein's field equations with constant $n$. These two solutions are derived from two roots of a quadratic equation. The purpose of presenting the solar system data is only to show that there is no discontinuity like the MOND function. One should not look for precision in Table~\ref{tab:Table2} because it is based on circular orbit approximation. It is sufficient to note that $n=1$ for flat Minkowski metric is recovered at small distances.

We can also apply these equations of stellar motion to rotation curves of M31 \cite{52} and NGC3198 \cite{53}. The results obtained from Eqs.~\eqref{1.36},~\eqref{1.33} and~\eqref{1.17} are shown in Tables~\ref{tab:Table3} and~\ref{tab:Table4}. Computed values are based on the stellar mass at the galactic center, which is $1.4\times10^{11}M_\odot$ for M31 and $5.0\times10^{9}M_\odot$ for NGC3198.  

From Eq.~\eqref{1.26}, we can see that $n$ is a ratio of Newtonian gravitational acceleration to the measured acceleration which is $1$ for flat Minkowski metric. From Eq.~\eqref{1.35} we get the same relation for circular orbits.
\begin{align}\label{1.39}
n=\frac{\mu/r^2}{v^2/r}.
\end{align}
Substitution of $n$ in Eq.~\eqref{1.17} gives
\begin{align}\label{1.40}
d\tau^2=\left(1-\frac{\mu}{rc^2}\right)dt^2.
\end{align}
Therefore metric~\eqref{1.40} becomes singular for the limiting radius
\begin{align}\label{1.41}
r_l=\frac{\mu}{c^2}.
\end{align}
This is the same expression which we derived earlier \cite{44}for a black hole. 

\section{Conclusion}
We have presented derivation for the deflection of light from fundamentals by introducing vectors. Here we can relate the additional component of acceleration with the rotation of the velocity vector which causes the curvature of the trajectory. We have distinguished the cartesian curvilinear acceleration from the polar conic acceleration and explained why they are not equal even though they are derived from the same velocity vector. We have derived expression for the Lorentz invariant acceleration. We have presented a theory of rotation curves of galaxies which is based on the second solution of Einstein's field equations which yields much better results than the earlier one based on the first solution with constant $n$ \cite{73}. Deviation factor $n$ appears in the expression for acceleration as well as the modified Kepler's third law which now yeilds correct orbital periods for the stars of galaxies. Deviation factor $n$ plays the same role as the MOND function in the expression for acceleration. This kind of solution cannot be obtained in general relativity because of the weak field approximation, which is a different way of introducing deviation to the flat Minkowski metric. 

\section{Acknowledgment}
Author is grateful to Robert Low, Gerard t'Hooft, Stam Nicolis, Christopher Eltschka, Bruce Rout and Thiago C. Junqueira for useful discussion and comments.

\pdfbookmark{References}{Ref}
%\vspace{5mm}
\bibliographystyle{amsalpha}

\begin{thebibliography}{1}
\bibitem {44} V. H. Zaveri, \textit{Periodic relativity: basic framework of the theory}. Gen. Relativ. Gravit. \textbf{42} 6, 1345--1374, (2010), doi: \href{http://dx.doi.org/10.1007/s10714-009-0908-5}{10.1007/s10714-009-0908-5}.
\href{https://www.researchgate.net/publication/225712819}{https://www.researchgate.net/publication/225712819}. \href{http://arxiv.org/abs/0707.4539v9}{arXiv:0707.4539v9}[physics.gen-ph].
\bibitem {77} V. H. Zaveri, \textit{Quarkonium and hydrogen spectra with spin dependent relativistic wave equation}, Pramana - J. Phys. \textbf{75}(4), 579--598, (2010). 
doi:\href{http://dx.doi.org/10.1007/s12043-010-0140-6} {10.1007/s12043-010-0140-6}.
%doi: 10.1007/s12043-010-0140-6.
%{doi:10.1007/s12043-010-0140-6}
\href{http://arxiv.org/abs/0707.2431v7}{arXiv: 0707.2431v7}.
\bibitem {20} V. H. Zaveri, \textit{Orbital period derivative of a binary system using an exact orbital energy equation}, \href{http://arxiv.org/abs/0707.4544v3}{arXiv:0707.4544v3}[physics.gen-ph].
\bibitem {50} V. H. Zaveri, \textit{Periodic quantum gravity and cosmology}. submitted, Grav. Cosmol. (May 2014).
\href{https://www.researchgate.net/publication/262492889}{https://www.researchgate.net/publication/262492889}
\bibitem {104} V. H. Zaveri, 2012$a$ Consciousness and energy. In: \textit{The Big Bang: Theory, Assumptions and Problems}, eds. J.R. O'Connell, \& A.L. Hale, 275--284. New York: Nova Science Publishers.
\href{https://www.novapublishers.com/catalog/product_info.php?products_id=21109}{https://www.novapublishers.com}
\href{https://www.researchgate.net/publication/262494186}{https://www.researchgate.net/publication/262494186}
\bibitem {55} V. H. Zaveri, \textit{Unified field of consciousness}. submitted,\\ (2014).\\
\href{https://www.researchgate.net/publication/266618399}{https://www.researchgate.net/publication/266618399}
\bibitem {73} V. H. Zaveri, \textit{Proper time in rotation curves of galaxies and accelerated expansion of universe
}, \href{http://arxiv.org/abs/0805.2233v4}{arXiv:0805.2233v4}[physics.gen-ph].
\bibitem {19}  X.-X. Xue, H.-W. Rix, G. Zhao, P. R. Fiorentin, T. Naab, M. Steinmetz, F. C. van den Bosch, T. C. Beers, Y. S. Lee, E. F. Bell, C. Rockosi, B. Yanny, H. Newberg, R. Wilhelm, X. Kang, M. C. Smith and D. P. Schneider, \textit{The Milky Way's Rotation Curve to 60 kpc and an Estimate of the Dark Matter Halo Mass from Kinematics of ~2500 SDSS Blue Horizontal Branch Stars}, Astrophys. J. \textbf{684}, 1143--1158, (2008). arXiv:0801.1232v3[astro-ph].
\bibitem {41} T. Naab, P.H. Johansson, J.P. Ostriker, G. Efstathiou, \textit{Formation of Early-Type Galaxies from Cosmological Initial Conditions}, Astrophys. J. \textbf{658}, 710N, (2007).
\bibitem {26}  F. Eisenhauer, R. Schoedel, R. Genzel, T. Ott, M. Tecza, R. Abuter, A. Eckart and T. Alexander, \textit{A Geometric Determination of the Distance to the Galactic Center}, Astrophys. J. Lett. \textbf{597}, L121-L124, (2003).
\bibitem {18} X. Wu, B. Famaey, G. Gentile, H. Perets and HS. Zhao, \textit{Milky Way potentials in CDM and MOND. Is the Large Magellanic Cloud on a bound orbit?}, Mon. Not. R. Astron. Soc. \textbf{386} 4, 2199--2208, (2008).  arXiv:0803.0977v1[astro-ph].
\bibitem {42} A. C. Robin,	C. Reyle,	S. Derriere, S. Picaud, \textit{A synthetic view on structure and evolution of the Milky Way}, Astron. Astrophys. \textbf{409}, 523, (2003).
\bibitem {52} E. Corbelli, S. Lorenzoni, R. Walterbos, R. Braun and D. Thilker, \textit{A wide-field H I mosaic of Messier 31 - II. The disk warp, rotation, and the dark matter halo}, Astron. Astrophys. 511, A89, (2010).
\bibitem {53} K.G. Begeman, \textit{H I rotation curves of spiral galaxies. I - NGC 3198}, Astron. Astrophys. 223, 47-60, (1989). 
\end{thebibliography}

\end{document}